\documentclass{aa}
\usepackage[dvips]{graphicx}
\usepackage{psfig}

\newcommand{\lapp}{\mbox{\raisebox{-0.3em}{$\stackrel{\textstyle <}{\sim}$}}}

\begin{document}

\thesaurus{}

\title{Small-scale structures in Compact Steep-Spectrum and 
GHz-Peaked-Spectrum radio sources}

\author{ S. Jeyakumar \inst{1} \and D.J. Saikia \inst{1} \and
A. Pramesh Rao \inst{1} \and  V. Balasubramanian \inst{2}  }

\titlerunning{Small-scale structures in CSS and GPS radio sources}

\institute{
National Centre for Radio Astrophysics, 
Tata Institute of Fundamental Research, Post Bag No. 3,\\
Ganeshkhind, Pune 411 007, India 
\and
Radio Astronomy Centre, Tata Institute of Fundamental Research, 
Post Box No. 8, Udhagamandalam 643 001, India 
}

\date{Received 0000; Accepted 0000}

\maketitle

\begin{abstract}
We determine the small-scale structure 
for a large sample of Compact Steep Spectrum (CSS) and Gigahertz Peaked Spectrum 
(GPS) sources and a sample of larger radio sources at 327 MHz using the
technique of Inter-Planetary Scintillation (IPS) with the Ooty Radio Telescope.
Our observations provide structural information on 
angular scales ranging from about 50 to a few hundred milliarcsec.
We estimate the size and flux density of the scintillating components from 
the IPS observations.  The 
fraction of flux density of the hotspots estimated from both the IPS
observations as well as from interferometric observations for larger sources
from the literature exhibits no significant dependence
on either the radio luminosity or linear size for the high-luminosity sources.
We examine the collimation of radio jets using the sizes of hotspots
from both IPS and interferometric observations. The hotspot 
sizes do not exhibit a significant dependence on the overall linear size
for the larger sources, although the CSS and GPS sources were earlier found
to evolve in a self-similar way. The IPS observations show evidence of 
larger-scale structures compared to the known VLBI-scale structures 
in 8 CSS and GPS radio sources.
We discuss the origin of these structures. We also examine the
spectra of compact components in GPS sources using both IPS and interferometric 
measurements, and attempt to distinguish between different processes 
for explaining their low-frequency spectra.
\end{abstract}


\section{Introduction }
Compact steep-spectrum sources (CSSs) are defined to be less than about
20 kpc in size (H$_o = 50$ km
s$^{-1}$ Mpc$^{-1}$ and $q_0 = 0$), and have a steep high-frequency radio
spectrum ($\alpha \geq 0.5$),
where S$\propto \nu^{-\alpha}$. This class of objects includes the more compact 
Gigahertz-Peaked Spectrum (GPS) sources with a turnover in their spectrum 
at frequencies in the GHz range. The larger CSS sources also sometimes exhibit 
a flattening or  turn-over in 
the integrated spectrum, but the turnover frequency is usually in the range of 
a few hundred MHz. High-resolution radio observations of CSS and GPS sources
(cf. Fanti et al. 1990; Sanghera et al. 1995 and references therein) reveal
a variety of structures reminiscent of those seen in the more extended
sources. The majority of sources have a double-lobed structure, often with a
nuclear or core component. Only a small fraction of sources appear to have
a highly distorted or complex structure. 

A number of possible explanations have been explored to understand the nature of
CSS and GPS objects (cf. O'Dea 1998). It has been argued that only a small
fraction are likely to appear small because of projection effects (Fanti et al. 1990).
Although some may be distorted and confined to small dimensions by a dense interstellar
medium in the host galaxy (cf. Wilkinson et al. 1991a,b; Carvalho 1994, 1998), 
there appears to be a consensus that their small sizes are 
largely because they are young sources seen at an early stage of 
their evolution (Carvalho 1985; Fanti  et al. 1995; Begelman 1996; 
Readhead et al. 1996a,b). The GPS sources with a
double-lobed structure, the compact doubles (CD), have been proposed to be 
miniature versions of the classical Fanaroff-Riley class II sources, and 
have been suggested to evolve from a CD to a CSS and then on to a larger
FRII source (Phillips \& Mutel 1981; Hodges et al. 1984; Mutel \& Phillips
1988; Carvalho 1985). 

In this paper we present the results of an Inter-Planetary Scintillation (IPS)
study of a large sample of CSS and GPS radio sources at 327 MHz using the 
Ooty Radio Telescope (ORT). These observations have been 
made with the objectives of
(i) determining the sizes and prominence of hotspots to investigate the
collimation of radio jets and the evolution of these sources; (ii) investigating
the existence of radio halos at a low frequency on the scale of hundreds of 
milliarcsec in 
addition to the more compact structure, suggestive of earlier periods of
activity; and
(iii) determining the low-frequency spectra of compact components in an
attempt to distinguish between different processes to
explain the low-frequency turnover in the spectra of GPS sources. We
describe the IPS observations, the sample of sources observed by this
technique and the results of the IPS observations in Section 2. The dependence
of the prominence of hotspots from both IPS and interferometric measurements
on radio luminosity and linear size, as well as
the results on the collimation of radio jets and possible constraints
on evolution of radio sources are presented in Section 3. The evidence for
the existence of possible large-scale structures on the scale of hundreds of mas
is presented in Section 4. In Section 5, we
present the spectra of 5 GPS sources where we have attempted to determine
the spectra of the dominant component from IPS and interferometric 
observations, and discuss possible reasons for their spectral shape.

\section{IPS Observations}

A compact radio source or component seen through the solar wind 
exhibits scintillations due to electron density fluctuations in the 
interplanetary medium. The power spectrum of the fluctuations depends on
the conditions of the solar wind, and the size and structure of the compact
component. Using a suitable model of the solar wind, IPS observations 
can be used to estimate the angular size of the scintillating component
from the power spectrum, and the scintillation index, m, which is defined
as $\delta$S/S. Here $\delta$S is the rms of the flux density variations and
S is the total flux density of the source 
(e.g. Little \&  Hewish 1966, 1968; Cohen et al. 1967; 
Rao et al. 1974). By comparing m$_{obs}$ for a given 
source with a point source calibrator, m$_{cal}$, we can estimate $\mu$, the
fraction of flux density in the scintillating component. 
Generally the scintillation index, m,
decreases with source size and becomes non-detectable when the source size
is so large that the signal to noise ratio is poor. This limits the 
detectable IPS size to about 0.$^{\prime\prime}$4 at 327 MHz, which we will
refer to as the IPS cut-off size. Also for the typical solar wind parameters, 
and the observing frequency of 327 MHz, the minimum size of the
components which can be estimated is about 50 mas
(Manoharan \& Ananthakrishnan 1990; Gothoskar \& Rao 1999).
In a typical radio source the compact scintillating components could be
the hotspots at the outer edges of the lobes, the nuclear or core
components or prominent knots in a jet. If the separation of the compact
components, such as the hotspots, is smaller than the IPS cut-off size, the source scintillates 
as a single source. On the other hand, if the scintillating components in the 
source are separated by larger than the IPS cut-off size, the components
scintillate independently and the parameters estimated from IPS observations 
are the weighted average values of the different scintillating components. 

\setcounter{table}{0}
\begin{table*}
\caption{Properties of the sample of largely CSS and GPS sources}
\label{vlbiinfo1}
\begin{tabular}{ l l l l l l l c c l l}
\hline
IAU &   Alt.name &  ID & z  & Sp & $S_{327}$ & $P_5$  & LAS & Size & Morph & Ref  \\
    &  &  &  & & Jy & WHz$^{-1}$sr$^{-1}$& \arcsec & kpc &  & \\
\hline
0019$-$000 &  4C +00.02 &   G & 0.305  &GPS   &  2.6 & 25.67  &  0.06  &0.36  &AD  &1,32\\
0023$-$263 &  OB $-$238 &   G & 0.322  &    & 19.3 & 26.21  &  0.55  &3.42  &D  &57\\
  0026+346 &         S4 &   G & 0.600  &GPS   &  1.5* & 26.32  &  0.03  &0.266  &T  &73\\
  0039+230 &        PKS &  EF &        &GPS   &  0.5 &       &       &       &T?/Cpx  &80\\
0044$-$056 & 4C $-$05.03 &   Q & 1.869  &    &  1.8 & 26.75  &     1  &12.8  &D?  &62,71,78\\
0114$-$211 &  OC $-$224 &   G &        &    & 11.4 &       &   0.6  &       &AT  &9,84,85\\
  0121+240 &        PKS &     &        &    &  2.1 &       &   2.4  &       &D  &81\\
  0127+233 &      3C 43 &   Q & 1.459  &    &  8.4 & 27.30  &   2.6  &31.5  &MT  &2,6,28,31\\
  0134+329 &      3C 48 &   Q & 0.367  &    & 44.0 & 26.52  &     1  &6.76  &CJ/Cpx  &16,22,31,34,49\\
  0138+136 &      3C 49 &   G & 0.621  &    &  7.6 & 26.31  &  0.92  &8.28  &T  &2,28,31\\
  0144+209 &        PKS &  EF &        &GPS   &  3.1 &       &       &       &-  &-\\
  0201+113 &    OD +101 &   Q & 3.610  &GPS   &  0.3 & 27.94  &   3.7  &51.2  &AD  &1,15,32\\
  0218+357 &         S4 &   G & 0.685  &GPS   &  1.9 & 26.29  &  0.36  &3.39  &See notes  &38,39,40\\
  0221+276 &      3C 67 &   G & 0.310  &    &  8.7 & 25.62  &  2.47  & 15  &T  &2,13,16,28,30,31\\
  0223+341 &  4C +34.07 &   Q &        &    &  3.6* &       &   1.1  &       &T  &31,41\\
  0319+121 &    OE +131 &   Q & 2.662  &FSC   &  2.6 & 27.85  &  0.06  &0.807  &See notes  &41,42,86\\
  0320+053 &  4C +05.14 &   G & 0.575  &    &  6.9 & 26.24  &  0.03  &0.26  &S  &4,24\\
  0345+337 &    3C 93.1 &   G & 0.243  &    &  6.3 & 25.30  &  0.48  &2.46  &Cpx  &30,31,41\\
  0358+004 &      3C 99 &   G & 0.426  &    &  5.4 & 25.69  &   4.4  &32.5  &AT  &43,44\\
  0400+258 &    CTD 026 &   Q & 2.109  &GPS   &  1.3 & 27.61  &   5.1  &66.5  &AD  &15,42\\
  0422+178 &  4C +17.25 &     &        &    &  2.0 &       &  0.76  &       &D  &81\\
  0428+205 &    OF +247 &   G & 0.219  &GPS   &  2.8 & 25.64  &  0.21  &0.998  &T  &31,41,66\\
  0445+097 &  4C +09.17 &   Q & 2.110  &    &  2.9 & 27.27  &   1.3  &16.9  &D?  &71,78\\
  0457+024 &    OF +097 &   Q & 2.384  &GPS   &  0.2* & 27.95  & 0.012  &0.159  &T  &80\\
  0500+019 &    OG +003 &   Q & 1.000  &GPS   &  0.5 & 27.01  & 0.015  &0.164  &T?  &1,12,20\\
  0518+165 &      3C138 &   Q & 0.759  &    & 19.8 & 27.12  &  0.62  &6.1  &T  &16,28,31,37,46\\
  0528+134 &    OG +147 &   Q & 2.060  &GPS   &  0.9 & 27.78  &     1  & 13  &AD  &47,57\\
  0531+194 &        PKS &   G &        &    & 15.1* &       &   0.5  &       &S  &57\\
0624$-$058 &      3C161 &   Q &        &    & 51.2 &       &   1.3  &       &CJ?  &33,55\\
  0658+380 &      3C173 &   G & 1.035  &    &  6.5 & 26.65  &  2.46  &27.1  &D  &30,31\\
  0711+356 &         S4 &   Q & 1.620  &GPS   &  0.6* & 27.38  &  13.5  &168  &T  &38,42,51,52,54\\
  0738+313 &    OI +363 &   Q & 0.631  &FSC   &  1.0 & 26.56  &  66.3  &601  &T  &12,42,86\\
  0740+380 &      3C186 &   Q & 1.063  &    &  7.7 & 26.49  &   2.2  &24.5  &T  &28,29,56,58\\
0741$-$063 & 4C $-$06.18 &   Q &        &GPS   & 10.2 &       &       &       &-  &-\\
  0742+103 &        PKS &     &        &GPS   &  1.2 &       & 0.012  &       &CJ?  &12,15\\
0743$-$006 & 4C $-$00.28 &   Q & 0.994  &GPS   &  0.3 & 26.77  & 0.011  &0.12  &CJ?  &12,15\\
  0751+298 &  4C +29.27 &   Q & 2.106  &    &  1.4 & 26.86  &   1.3  &16.9  &D?  &71,78\\
  0752+342 &        TXS &   G &        &    &  1.2 &       & 0.013  &       &S  &65\\
  0758+143 &      3C190 &   Q & 1.970  &    &  9.1 & 27.58  &   2.6  &33.5  &MT  &16,28,29,33\\
  0802+103 &      3C191 &   Q & 1.956  &GPS   &  7.9 & 27.53  &   4.6  &59.2  &T  &2,33,34,58,78\\
  0843+136 &  4C +13.39 &   Q & 1.877  &    &  1.5 & 26.82  &   1.8  & 23  &D  &71,78\\
  0848+155 &     OJ 180 &   Q & 2.007  &    &  1.3* & 27.12  &   0.5  &6.46  &D  &71,78\\
  0858+292 &    3C213.1 &   G & 0.194  &    &  5.6 & 25.05  &  23.2  &101  &D  &30,31,79\\
  0904+039 &        PKS &     &        &GPS   &  0.9* &       &       &       &-  &-\\
  0914+114 &        PKS &   G &        &GPS   &  2.4* &       &       &       &-  &-\\
0941$-$080 &        PKS &   G & 0.228  &GPS   &  3.9 & 25.37  &  0.05  &0.245  &D  &11,19\\
  0941+261 &     OK 270 &   Q & 2.908  &    &  2.2* & 27.56  &   1.5  &20.4  &MT  &71,78\\
  1005+077 &      3C237 &   G & 0.877  &    & 19.0 & 27.07  &   1.3  &13.5  &T  &16,28,31,33,36\\
  1019+222 &      3C241 &   G & 1.617  &    &  7.6 & 27.24  &  0.81  &10.1  &MT  &2,28,30,31,33,36\\
  1045+019 &        PKS &     &        &FSC   &  0.4 &       &   3.1  &       &MT  &32,80\\
  1117+146 &  4C +14.41 &   G & 0.362  &GPS   &  3.7 & 25.78  & 0.087  &0.583  &T  &2,60,69,87\\
  1122+195 &      3C258 &   G & 0.165  &    &  2.5 & 24.62  &    60  &230  &D  &2,23,30,77\\
1127$-$145 &        PKS &   Q & 1.187  &GPS   &  5.7 & 27.53  & 0.063  &0.724  &D/CJ?  &48,61,63,64\\
1143$-$245 &        PKS &   Q & 1.950  &GPS   &  0.2 & 27.60  & 0.007  &0.09  &D/CJ?  &12,19\\
1148$-$171 &        PKS &   Q & 1.751  &GPS   &  0.8* & 27.23  &       &       &-  &-\\
  1153+317 &  4C +31.38 &   Q & 0.418  &    &  7.2 & 25.92  &  0.89  &6.5  &T  &31,35\\
  1210+134 &  4C +13.46 &   Q & 1.141  &    &  2.4 & 26.72  &   0.5  &5.68  &S  &78\\
\hline
\end{tabular}
\end{table*}
\setcounter{table}{0}
\begin{table*}
\caption{Contd.}
\label{vlbiinfo1c}
\begin{tabular}{ l l l l l l l c c l l}
\hline
IAU &   Alt.name &  ID & z  & Sp & $S_{327}$ & $P_5$  & LAS & Size & Morph & Ref  \\
    &  &  &  & & Jy & WHz$^{-1}$sr$^{-1}$ & \arcsec & kpc &  & \\
\hline
  1221+113 &        TXS &   Q & 1.758  &    &  1.0* & 26.60  &   1.3  &16.4  &D/T?  &71,78\\
1237$-$101 &  ON $-$162 &   Q & 0.753  &FSC   &  2.0 & 26.52  &    17  &167  &-  &57\\
1245$-$197 &        PKS &   Q & 1.275  &GPS   &  8.1 & 27.49  &   0.3  &3.52  &AD/CJ?  &84\\
  1341+144 &  4C +14.49 &     &        &    &  4.5 &       &  0.09  &       &D/CJ?  &65,81\\
  1345+125 &  4C +12.50 &   G & 0.120  &GPS   &  7.9 & 25.19  & 0.084  &0.248  &MT?  &8,12,63\\
1354$-$174 &        PKS &   G & 3.147  &GPS   &  0.7* & 28.24  & 0.032  &0.438  &AD/AMT?  &10\\
  1416+067 &      3C298 &   Q & 1.436  &    & 32.9 & 27.55  &  1.49  & 18  &T  &16,28,30,31,33,78\\
1433$-$040 &        PKS &   G &        &GPS   &  0.7* &       &       &       &-  &-\\
  1442+101 &     OQ 172 &   Q & 3.522  &GPS   &  1.8 & 28.47  &  0.02  &0.276  &CJ?  &3,41\\
  1456+092 &     OQ 095 &   Q & 1.991  &    &  1.8* & 27.03  &     2  &25.8  &D?  &62,78\\
  1518+047 &  4C +04.51 &   G & 1.296  &GPS   &  2.6 & 27.45  &  0.14  &1.65  &D  &15,19,66,67,68\\
1524$-$136 &        PKS &   Q & 1.687  &    &  6.1 & 27.49  & 0.538  &6.74  &MT  &9,57,59,84\\
  1543+005 &        PKS &   G & 0.550  &GPS   &  2.3* & 26.19  & 0.007  &0.0594  &T?  &80\\
1548$-$302 &        PKS &     &        &GPS   &  1.7* &       &       &       &-  &-\\
1601$-$222 &        PKS &   G &        &    &  0.9* &       &       &       &-  &-\\
2008$-$068 &        PKS &   G &        &GPS   &  1.2 &       &  0.03  &       &MT?  &57,80\\
2044$-$027 &      3C422 &   Q & 0.942  &    &  6.8 & 26.67  &     1  &10.7  &AD/Cpx  &19,45\\
2053$-$201 &        PKS &   G & 0.156  &GPS   &  6.5 & 24.94  &       &       &-  &-\\
2121$-$014 &        PKS &     &        &GPS   &  1.9* &       &       &       &-  &-\\
2126$-$185 &        PKS &   Q & 0.680  &GPS   &  1.0* & 26.41  &       &       &-  &-\\
  2128+048 &        PKS &   G & 0.990  &GPS   &  3.4 & 27.13  & 0.035  &0.38  &T  &1,12,19\\
  2134+004 &        PKS &   Q & 1.932  &GPS   &  0.9 & 28.11  &   2.8  & 36  &AD  &20,21,42,63\\
2143$-$156 &        PKS &   Q & 0.701  &    &  0.6* & 26.47  &       &       &-  &-\\
2146$-$133 &        PKS &   Q & 1.800  &    &  5.0 & 27.32  &   3.5  &44.4  &T  &71,78\\
  2147+145 &        PKS &     &        &    &  6.5 &       & 0.022  &       &D?  &27,65,81\\
2149$-$306 &        PKS &   Q & 2.345  &FSC   &  0.8* & 27.59  &     3  &39.7  &-  &57\\
  2149+056 &        PKS &   G & 0.740  &GPS   &  0.2 & 26.32  &0.0035  &0.0341  &AD?  &80\\
2153$-$119 &       PMNJ &     &        &GPS   &  0.5 &       &0.0017  &       &D?  &80\\
2154$-$183 &        PKS &   Q & 1.423  &GPS   &  3.0* & 27.23  &       &       &-  &-\\
  2158+101 &  4C +10.67 &   Q & 1.725  &    &  1.7 & 26.53  &   0.8  &10.1  &D?  &71,78\\
  2210+016 &  4C +01.69 &   G & 0.431  &GPS   &  4.3 & 25.98  & 0.075  &0.558  &MT  &12,19\\
  2222+051 &  4C +05.84 &   Q & 2.323  &    &  3.2 & 27.33  &     3  &39.7  &T  &71,78\\
  2223+210 &        PKS &   Q & 1.959  &FSC   &  3.3 & 27.52  &   5.3  &68.2  &T  &42,70,78\\
  2230+114 &  4C +11.69 &   Q & 1.037  &FSC   &  7.5 & 27.34  &   2.6  &28.7  &T  &31,53,54,57,69,72\\
2243$-$123 &        PKS &   Q & 0.630  &FSC   &  1.6* & 26.59  &   3.9  &35.3  &OS  &57\\
  2247+140 &  4C +14.82 &   Q & 0.237  &    &  3.7 & 25.39  &   0.2  &1.01  &D?  &31,35\\
  2252+129 &      3C455 &   Q & 0.543  &    &  8.8 & 26.14  &  3.12  &26.3  &D  &30,31,78\\
2322$-$040 &        PKS &   G &        &GPS   &  0.7 &       &       &       &-  &-\\
  2338+132 &  4C +13.88 &     &        &    &  6.5 &       &  0.98  &       &D  &81\\
2345$-$167 &        PKS &   Q & 0.576  &FSC   &  1.1* & 26.62  &   3.8  & 33  &-  &7,48,57,76\\
  2345+061 &  4C +06.76 &   Q & 1.546  &    &  1.9 & 26.68  &  0.98  & 12  &MT/Cpx?  &62,78\\
  2353+154 &        PKS &   Q & 1.801  &    &  2.6 & 27.10  &  0.05  &0.634  &S  &78\\
\hline
\end{tabular}
\\
Notes: 0218+357: This source is gravitationaly lensed (O'Dea et al. 1992; Patnaik et al. 1995). The
LAS quoted here is the separation between the two lensed images (Polatidis et al. 1995).
0319+121: This spectrum of the source is flat (c.f. Dallacasa et al. 1995; Saikia et al. 1998).
The LAS is estimated using the extended emission seen by Murphy et al. (1993).
\end{table*}

\setcounter{table}{1}
\begin{table*}
\caption{Properties of the sample of larger sources}
\label{vlbiinfo2}
\begin{tabular}{ l l l l l l l c c l l}
\hline
IAU &   Alt.name &  ID & z  & Sp & $S_{327}$ & $P_5$  & LAS & Size & Morph & Ref  \\
    &  &  &  & & Jy & WHz$^{-1}$sr$^{-1}$ & \arcsec & kpc &  & \\
\hline
0038$-$019 & 4C$-$02.04 &   Q & 1.690  &    &  3.7* & 27.15  &  19.4  &243  &T  &62,78\\
  0109+176 &   4C+17.09 &   Q & 2.157  &    &  2.0* & 26.98  &  13.2  &173  &MT  &62,78\\
  0115+027 &   4C+02.04 &   Q & 0.672  &    &  5.6 & 26.14  &  13.1  &122  &T  &74,82\\
  0154+286 &       3C55 &   G & 0.240  &    &  9.5 & 25.18  &    69  &351  &T  &14\\
0222$-$008 & 4C$-$00.12 &   Q & 0.687  &    &  2.9 & 26.03  &  13.5  &127  &T  &74\\
0232$-$042 & 4C$-$04.06 &   Q & 1.436  &    &  4.5 & 27.01  &    13  &157  &MT  &74\\
0350$-$073 &       3C94 &   Q & 0.962  &    & 11.9 & 26.75  &  41.5  &446  &D  &50\\
  0404+035 &      3C105 &   G & 0.089  &    & 13.1 & 24.82  &   335  &763  &T  &18\\
0736$-$019 &      3C185 &   G & 1.033  &    &  3.8 & 26.47  &   8.9  &98.1  &D  &88\\
  0850+140 &      3C208 &   Q & 1.110  &    &  9.6 & 26.81  &    14  &158  &T  &83\\
  0855+285 &      3C210 &   G & 1.169  &    &  7.0 & 26.79  &  17.5  &200  &D  &23\\
  0855+143 &      3C212 &   Q & 1.048  &    &  8.8 & 26.85  &    10  &111  &T  &5\\
  0941+100 &      3C226 &   G & 0.823  &    &  9.7 & 26.44  &    31  &315  &T  &17,75\\
  1012+022 &   4C+02.30 &   Q & 1.374  &    &  2.0 & 26.52  &   8.5  &102  &MT  &26\\
  1015+277 &   4C+27.21 &   Q & 0.469  &    &  3.4 & 25.69  &    19  &148  &D  &25\\
  1022+194 &   4C+19.34 &   Q & 0.828  &    &  1.3 & 26.04  &    44  &448  &T  &74\\
  1023+067 &      3C243 &   Q & 1.699  &    &  4.6 & 26.88  &  13.3  &167  &T  &78\\
1136$-$135 &        PKS &   Q & 0.557  &    & 13.4 & 26.49  &  15.8  &135  &T  &82\\
\hline
\end{tabular}
\end{table*}

\begin{table*}
\caption{Key to the references for Tables 1 and 2}
\begin{tabular}{l l l l l l }
  1 & \cite{  1}  &   31 & \cite{ 19}  &   61 & \cite{ 59}   \\
  2 & \cite{ 10}  &   32 & \cite{  2}  &   62 & \cite{  6}   \\
  3 & \cite{100}  &   33 & \cite{ 21}  &   63 & \cite{ 60}   \\
  4 & \cite{101}  &   34 & \cite{ 22}  &   64 & \cite{ 61}   \\
  5 & Laing pvt. comm. &   35 & \cite{ 23}  &   65 & \cite{ 62}   \\
  6 & \cite{ 20}  &   36 & \cite{ 25}  &   66 & \cite{ 64}   \\
  7 & \cite{104}  &   37 & \cite{ 26}  &   67 & \cite{ 65}   \\
  8 & \cite{105}  &   38 & \cite{ 29}  &   68 & \cite{ 66}   \\
  9 & \cite{106}  &   39 & \cite{ 30}  &   69 & \cite{ 56}   \\
 10 & \cite{107}  &   40 & \cite{ 31}  &   70 & \cite{ 69}   \\
 11 & \cite{108}  &   41 & \cite{ 32}  &   71 & \cite{  7}   \\
 12 & \cite{109}  &   42 & \cite{ 33}  &   72 & \cite{ 70}   \\
 13 & \cite{ 11}  &   43 & \cite{ 36}  &   73 & \cite{  5}   \\
 14 & \cite{111}  &   44 & \cite{ 37}  &   74 & \cite{ 75}   \\
 15 & \cite{112}  &   45 & \cite{ 39}  &   75 & \cite{ 76}   \\
 16 & \cite{ 12}  &   46 & \cite{ 40}  &   76 & \cite{ 78}   \\
 17 & \cite{121}  &   47 & \cite{ 42}  &   77 & \cite{ 35}   \\
 18 & \cite{129}  &   48 & \cite{ 43}  &   78 & \cite{ 7A}   \\
 19 & \cite{131}  &   49 & \cite{ 34}  &   79 & \cite{139}   \\
 20 & \cite{110}  &   50 & \cite{145}  &   80 & \cite{140}   \\
 21 & \cite{132}  &   51 & \cite{ 48}  &   81 & \cite{  8}   \\
 22 & \cite{133}  &   52 & \cite{ 49}  &   82 & \cite{ 81}   \\
 23 & \cite{134}  &   53 & \cite{ 72}  &   83 & \cite{ 85}   \\
 24 & \cite{ 15}  &   54 & \cite{ 50}  &   84 & \cite{  9}   \\
 25 & \cite{136}  &   55 & \cite{ 52}  &   85 & \cite{141}   \\
 26 & \cite{137}  &   56 & \cite{ 53}  &   86 & \cite{142}   \\
 27 & \cite{138}  &   57 & \cite{ 54}  &   87 & \cite{143}   \\
 28 & \cite{ 16}  &   58 & \cite{ 55}  &   88 & Saikia et al. in prep.  \\
 29 & \cite{ 17}  &   59 & \cite{ 68}  &     &         \\
 30 & \cite{ 18}  &   60 & \cite{ 58}  &     &         \\
\end{tabular}
\end{table*}

\subsection{ The sample of sources }
The sample of CSS and GPS sources for the IPS observations have been 
compiled from a number of papers (Gopal-Krishna et al. 1983; O'Dea et al. 1991;
Spoelstra et al. 1985; Savage et al. 1990; Cersosimo et al. 1994; Stanghellini
et al. 1990; Peacock \& Wall 1982; Sanghera et al. 1995), with the following
criteria: (i) declination $ |\delta| < 40^\circ$, the normal observable
range of the ORT; (ii) total flux density at 327 MHz,  $S_{327} >$ 200 mJy
so that the source can be observed with adequate signal to noise ratio; (iii) 
the ecliptic latitude of the source is $ < 30^\circ$ so that there is enough
scintillating power to estimate the parameters reliably. 
Many sources in our sample occur in more than one of the above lists and  they
have been counted only once in our list. 
Also, more recent estimates of sizes and redshifts have led to a number of
objects in our sample being larger than the canonical limit of 20 kpc. 
These have been retained in the list. A few of
the sources which were earlier identified as GPS sources appear to have a flat
radio spectrum. These have been marked separately in the Table and have not been
used in the statistical analyses.
Our final sample of largely compact sources consists of 100 objects, 48 of which 
are GPS sources. In addition, for comparison we have compiled a sample
of 19 larger-sized 3C and 4C sources of similar redshift and luminosity. All 
these sources have $S_{327} >$ 1 Jy, and their linear
sizes are larger than about 100 kpc for all but one source.

The basic radio and optical properties of the sample of largely CSS and GPS sources
and the sample of larger sources are 
listed in Tables 1 and 2 respectively, which are arranged as follows. Columns 1 and 2: IAU name
and an alternative name; column 3: optical identification where G denotes a galaxy,
Q a quasar and EF an empty field; 
column 4: redshift; column 5: identification of the object as a GPS source. Some of the
sources which were tentatively classified as GPS objects appear to have a flat radio 
spectrum. These have been indicated as an FSC or flat-spectrum core.
Column 6:  flux density at 327 MHz in Jy. 
The flux density values are interpolated from a least-squares 
fit to the data compiled from the literature, as described in Steppe et al. (1995). 
A $\star$  denotes an extrapolated value 
in which a measurement of the flux density was available at about 400 MHz. 
Column 7:  radio luminosity at 5 GHz in units of W Hz$^{-1}$ sr$^{-1}$; 
column 8: the largest angular size of the source (LAS) in arcsec; for sources with
weak extended emission, the LAS is defined  by the bright features on opposite sides.
Column 9: the projected linear size in kpc; column 10: structure of radio emission, 
where T denotes
a triple source, MT a misaligned triple where the supplement of the angle formed at the
core by the outer hotspots exceed 15$^\circ$, D  a double-lobed one, CJ for a core-jet 
structure, S for a 
single component and Cpx for a complex structure. A CJ source is one where there is one
dominant component with a jet-like extension rather than a second distinct lobe or hotspot.
The double-lobed and triple sources
where the flux density ratio of the outer components exceeds 10 are defined to
be asymmetric and are identified in this column as AD, AT or AMT. A flat-spectrum core
with extended emission on only one-side is indicated by the symbol OS. Any uncertainty in 
the information or classification is indicated by a question mark.
Column 11: references to radio structure. The key to the references are listed
in Table 3.

\subsection{Observations and analyses}

We have observed our sample of sources with the ORT 
at 327 MHz in the phase-switched  mode with a
bandwidth of 4 MHz. The  intensity variations  of a source
are sampled at a rate of 20ms.  
Each source was observed at a number of solar elongations, $\epsilon$, 
the observations at each elongation lasting about 15 - 20 minutes.
Our  observational procedure is similar
to the one described by Manoharan \& Ananthakrishnan(1990) and  Gothoskar \& Rao (1999).
We also monitored the off source region simultaneously in the total-power mode, to look
for any  interference, and edited the data accordingly. 
Since strong scattering occurs when $\epsilon <$ about 12$^\circ$,
observations were not made for lower values of $\epsilon$. 
Periods of large solar activity can also modify the average properties of the
solar wind and hence the observed power spectrum. Data during such
periods were also edited since we are primarily interested in the properties of
the sources (Manoharan 1993,1997). For each source,
the average power spectrum of the scintillations was obtained 
at each  elongation. From this spectrum, we estimate the scintillation index, m.
The typical error in the scintillation index is usually less than about 5 per cent. 
Variations in the interplanetary 
medium also leads to a scatter in the estimated value of 
$\mu$. The percentage uncertainty in the fraction of scintillating flux density, 
$\mu$, are listed in Table 4.

An additional source of uncertainity in $\mu$ is the  scintillations produced
by the confusing sources present in the beam. The confusion limit of the ORT
is about 1.5Jy. However, only a fraction of these sources have compact 
components which are likely to contribute to the scintillations. At low
frequencies the scintillating features would be almost entirely the hotspots
at the outer edges of the beams in the high-luminosity FRII sources.
Estimating the fraction of FRII radio sources in the B2 sample (Allington-Smith 1982) 
to be about 60 per cent, and assuming the fraction of flux density in the scintillating
hotspots to be about 50 per cent, yields a conservative estimate of the
scintillation confusion to be about 0.2 Jy. This affects significantly 
the interpretation of the scintillation visibility $\mu$ of the weaker sources.
The values for the weak sources with S$_{327} <$ 0.5 Jy have been 
listed in the tables, but have not been used further in the discussions.

The variation of scintillation index, m, with solar elongation  
for a point source is given by m$\propto(sin\epsilon)^{-\beta}$. 
The index $\beta$ is obtained by a least-squares fit to the
observed scintillation index of the IPS calibrator, at various elongations.
The compact radio source 1148$-$001 is the standard IPS calibrator used
at the ORT (cf. Venugopal et al. 1985), and observations of this source
during 1994$-$95 have been used to calibrate our observed scintillation indices.
The observed scintillation index for a given source can be smaller than the 
calibrator at the same elongation if either 
only a fraction $\mu$ of the total flux density is in the scintillating component, 
$m_{obs}(\epsilon) = \mu m_{cal}(\epsilon)$, or the scintillating component is extended. 
We apply a correction to $m_{obs}$, for the finite size of the scintillating component 
using the model of solar wind (Manoharan \& Ananthakrishnan 1990; 
Gothoskar \& Rao 1999). 
This correction could not be applied to 12 of the 91 sources with S$_{327} >$ 0.5 Jy 
in the sample of largely CSS and GPS radio sources (Table 1), and 6 of the 19 larger sources
(Table 2) since we could not estimate the size of the component reliably from our 
observed power spectrum.  

\setcounter{table}{3}
\begin{table}
\caption{Observed parameters of largely  CSS, GPS sources}
\label{ipstable1}
\begin{tabular}{l l c l l c }
\hline
IAU-name &  $N_s$ &$\theta_{1}\times\theta_{2}$ & PA &$N_{t}$ & $\mu$ \\
 &  & \arcsec $\times$ \arcsec & $^\circ$ &  & \\
\hline
0019$-$000 &    3 & $<$ 0.050 &   &    6 &  0.87 $\pm$0.06 \\
0023$-$263 &    2 & 0.162 &   &    4 &  0.36 $\pm$0.03 \\
  0026+346 &      &       &   &    1 &  0.38 $\pm$0.05 \\
  0039+230 &      &       &   &    6 &  0.88 $\pm$0.03 \\
0044$-$056 &    2 & 0.125 &   &    7 &  0.30 $\pm$0.04 \\
0114$-$211 &    6 & 0.350 $\times$  0.200 &      &     6 &  0.55 $\pm$0.03 \\
  0121+240 &    6 & 0.215 $\times$ $<$ 0.040 &    58 &    10 &  0.47 $\pm$0.03 \\
  0127+233 &    7 & 0.275 $\times$  0.125 &      &     8 &  0.32 $\pm$0.04 \\
  0134+329 &    6 & 0.225 &   &    6 &  0.31 $\pm$0.03 \\
  0138+136 &    4 & 0.081 &   &   10 &  0.39 $\pm$0.03 \\
  0144+209 &      &       &   &    4 &  0.08 $\pm$0.03 \\
  0201+113 &      &       &   &    4 &  0.87 $\pm$0.06 \\
  0218+357 &      &       &   &    4 &  0.20 $\pm$0.03 \\
  0221+276 &    4 & 0.206 &   &    4 &  0.34 $\pm$0.06 \\
  0223+341 &    4 & 0.194 &   &    5 &  0.27 $\pm$0.03 \\
  0319+121 &    3 & $<$ 0.050 &   &    4 &  0.58 $\pm$0.04 \\
  0320+053 &    2 & 0.100 &   &    6 &  0.73 $\pm$0.03 \\
  0345+337 &    6 & 0.183 &   &    6 &  0.36 $\pm$0.03 \\
  0358+004 &    7 & 0.218 &   &    7 &  0.44 $\pm$0.03 \\
  0400+258 &    1 & 0.125 &   &    3 &  0.60 $\pm$0.09 \\
  0422+178 &    2 & 0.125 &   &    2 &  0.42 $\pm$0.09 \\
  0428+205 &    2 & $<$ 0.050 &   &    2 &  0.54 $\pm$0.10 \\
  0445+097 &    2 & 0.125 &   &    5 &  0.27 $\pm$0.06 \\
  0457+024 &      &       &   &    2 &  1.12 $\pm$0.04 \\
  0500+019 &      &       &   &    2 &  0.77 $\pm$0.02 \\
  0518+165 &    2 & 0.112 &   &    3 &  0.13 $\pm$0.13 \\
  0528+134 &    2 & $<$ 0.050 &   &    3 &  0.55 $\pm$0.02 \\
  0531+194 &    2 & 0.115 &   &    2 &  0.29 $\pm$0.05 \\
0624$-$058 &    6 & 0.275 $\times$  0.200 &      &     6 &  0.22 $\pm$0.03 \\
  0658+380 &    5 & 0.138 &   &    6 &  0.27 $\pm$0.03 \\
  0711+356 &      &       &   &    4 &  0.65 $\pm$0.03 \\
  0738+313 &    2 & $<$ 0.050 &   &    3 &  0.42 $\pm$0.02 \\
  0740+380 &    3 & 0.120 &   &    3 &  0.18 $\pm$0.03 \\
0741$-$063 &    5 & 0.150 $\times$  0.075 &      &     5 &  0.54 $\pm$0.02 \\
  0742+103 &    3 & $<$ 0.050 &   &    3 &  0.82 $\pm$0.06 \\
0743$-$006 &      &       &   &    4 &  1.11 $\pm$0.05 \\
  0751+298 &      &       &   &    1 &  0.29 $\pm$0.05 \\
  0752+342 &    2 & 0.063 &   &    3 &  0.35 $\pm$0.04 \\
  0758+143 &    3 & 0.108 &   &    7 &  0.21 $\pm$0.04 \\
  0802+103 &    2 & 0.135 &   &    3 &  0.19 $\pm$0.05 \\
  0843+136 &    1 & 0.090 &   &    6 &  0.43 $\pm$0.03 \\
  0848+155 &    1 & $<$ 0.050 &   &    4 &  0.53 $\pm$0.04 \\
  0858+292 &      &       &   &    8 &  0.05 $\pm$0.05 \\
  0904+039 &    1 & 0.080 &   &    9 &  0.66 $\pm$0.03 \\
  0914+114 &    3 & 0.080 &   &    5 &  0.67 $\pm$0.04 \\
0941$-$080 &    6 & 0.089 &   &    8 &  0.61 $\pm$0.03 \\
  0941+261 &    4 & 0.082 &   &    6 &  0.31 $\pm$0.03 \\
  1005+077 &    2 & 0.110 &   &    4 &  0.08 $\pm$0.29 \\
  1019+222 &    2 & 0.138 &   &    3 &  0.53 $\pm$0.04 \\
  1045+019 &      &       &   &    4 &  0.62 $\pm$0.09 \\
  1117+146 &    6 & 0.088 &   &    8 &  0.73 $\pm$0.03 \\
  1122+195 &    9 & 0.235 $\times$  0.070 &   176 &    12 &  0.53 $\pm$0.03 \\
1127$-$145 &   15 & 0.350 $\times$ $<$ 0.050 &      &    17 &  0.49 $\pm$0.04 \\
1143$-$245 &      &       &   &    1 &  0.80 $\pm$0.05 \\
1148$-$171 &      &       &   &    3 &  0.24 $\pm$0.14 \\
  1153+317 &    4 & 0.200 $\times$  0.105 &      &     4 &  0.37 $\pm$0.04 \\
  1210+134 &    5 & 0.300 $\times$  0.085 &      &    14 &  0.38 $\pm$0.04 \\
\hline
\end{tabular}
\end{table}
\setcounter{table}{3}
\begin{table}
\caption{Contd.}
\label{ipstable1c}
\begin{tabular}{l l c l l c }
\hline
IAU-name &  $N_s$ &$\theta_{1}\times\theta_{2}$ & PA &$N_{t}$ & $\mu$ \\
 &  & \arcsec $\times$ \arcsec & $^\circ$ &  & \\
\hline
  1221+113 &      &       &   &    9 &  0.33 $\pm$0.06 \\
1237$-$101 &    2 & 0.213 &   &   10 &  0.56 $\pm$0.03 \\
1245$-$197 &    9 & 0.350 $\times$ $<$ 0.050 &      &    11 &  0.51 $\pm$0.03 \\
  1341+144 &    8 & 0.375 $\times$ $<$ 0.050 &      &    11 &  0.51 $\pm$0.03 \\
  1345+125 &    4 & 0.094 &   &    9 &  0.59 $\pm$0.03 \\
1354$-$174 &    2 & 0.188 &   &    9 &  1.10 $\pm$0.04 \\
  1416+067 &    9 & 0.305 $\times$  0.120 &   112 &    13 &  0.25 $\pm$0.05 \\
1433$-$040 &    2 & 0.133 &   &   10 &  1.00 $\pm$0.06 \\
  1442+101 &    1 & 0.150 &   &   10 &  0.67 $\pm$0.05 \\
  1456+092 &    1 & 0.125 &   &    8 &  0.35 $\pm$0.07 \\
  1518+047 &    8 & 0.325 $\times$  0.150 &      &    11 &  0.62 $\pm$0.04 \\
1524$-$136 &    5 & 0.100 &   &    8 &  0.47 $\pm$0.06 \\
  1543+005 &    8 & 0.325 $\times$  0.075 &      &     9 &  0.74 $\pm$0.04 \\
1601$-$222 &    1 & $<$ 0.050 &   &    3 &  0.32 $\pm$0.07 \\
2008$-$068 &    6 & 0.096 &   &   10 &  0.49 $\pm$0.05 \\
2044$-$027 &   14 & 0.245 $\times$  0.090 &    68 &    14 &  0.27 $\pm$0.03 \\
2053$-$201 &      &       &   &    3 &  0.07 $\pm$0.08 \\
2121$-$014 &    8 & 0.112 &   &   11 &  0.68 $\pm$0.04 \\
2126$-$185 &    1 & 0.100 &   &    4 &  0.79 $\pm$0.10 \\
  2128+048 &   10 & 0.100 &   &   10 &  0.62 $\pm$0.04 \\
  2134+004 &    9 & 0.175 $\times$ $<$ 0.050 &      &    13 &  0.54 $\pm$0.03 \\
2143$-$156 &    4 & 0.056 &   &    7 &  1.04 $\pm$0.05 \\
2146$-$133 &    6 & 0.136 &   &    7 &  0.23 $\pm$0.05 \\
  2147+145 &    5 & 0.095 &   &    5 &  0.49 $\pm$0.03 \\
2149$-$306 &    6 & 0.125 $\times$  0.075 &      &    10 &  0.57 $\pm$0.04 \\
  2149+056 &      &       &   &    3 &  0.69 $\pm$0.04 \\
2153$-$119 &      &       &   &    5 &  0.76 $\pm$0.06 \\
2154$-$183 &    3 & 0.108 &   &    4 &  0.61 $\pm$0.03 \\
  2158+101 &    1 & 0.100 &   &    7 &  0.23 $\pm$0.08 \\
  2210+016 &    9 & 0.106 &   &    9 &  0.72 $\pm$0.07 \\
  2222+051 &    7 & 0.225 $\times$  0.125 &      &     7 &  0.25 $\pm$0.06 \\
  2223+210 &    3 & 0.125 &   &    3 &  0.41 $\pm$0.07 \\
  2230+114 &    3 & 0.083 &   &    5 &  0.44 $\pm$0.06 \\
2243$-$123 &    4 & 0.081 &   &    7 &  0.72 $\pm$0.06 \\
  2247+140 &    6 & 0.325 $\times$  0.150 &      &     6 &  0.65 $\pm$0.04 \\
  2252+129 &    4 & 0.188 &   &    5 &  0.14 $\pm$0.04 \\
2322$-$040 &    1 & 0.125 &   &    6 &  0.50 $\pm$0.03 \\
  2338+132 &    3 & 0.112 &   &    3 &  0.45 $\pm$0.03 \\
2345$-$167 &    3 & 0.088 &   &    6 &  1.07 $\pm$0.03 \\
  2345+061 &      &       &   &    1 &  0.30 $\pm$0.05 \\
  2353+154 &    3 & 0.158 &   &    5 &  0.58 $\pm$0.04 \\
\hline
\end{tabular}
\end{table}
\setcounter{table}{4}
\begin{table}
\caption{Observed parameters of larger sources}
\label{ipstable2}
\begin{tabular}{l l c l l c }
\hline
IAU-name &  $N_s$ &$\theta_{1}\times\theta_{2}$ & PA &$N_{t}$ & $\mu$ \\
 &  & \arcsec $\times$ \arcsec & $^\circ$ &  & \\
\hline
0038$-$019 &      &       &   &    1 &  0.06 $\pm$0.05 \\
  0109+176 &      &       &   &    2 &  0.23 $\pm$0.04 \\
  0115+027 &    1 & 0.400 &   &    1 &  0.23 $\pm$0.05 \\
  0154+286 &      &       &   &    3 &  0.04 $\pm$0.03 \\
0222$-$008 &    2 & 0.125 &   &    3 &  0.33 $\pm$0.04 \\
0232$-$042 &    1 & 0.075 &   &    1 &  0.18 $\pm$0.05 \\
0350$-$073 &    2 & 0.200 &   &    2 &  0.18 $\pm$0.04 \\
  0404+035 &      &       &   &    1 &  0.04 $\pm$0.05 \\
0736$-$019 &    1 & 0.200 &   &    3 &  0.24 $\pm$0.08 \\
  0850+140 &    1 & 0.175 &   &    1 &  0.17 $\pm$0.05 \\
  0855+285 &    1 & 0.125 &   &    1 &  0.16 $\pm$0.05 \\
  0855+143 &    1 & 0.125 &   &    1 &  0.10 $\pm$0.05 \\
  0941+100 &    2 & $<$ 0.050 &   &    2 &  0.11 $\pm$0.02 \\
  1012+022 &    1 & 0.100 &   &    2 &  0.66 $\pm$0.10 \\
  1015+277 &    2 & 0.138 &   &    2 &  0.27 $\pm$0.04 \\
  1022+194 &      &       &   &    2 &  0.29 $\pm$0.02 \\
  1023+067 &    1 & 0.150 &   &    1 &  0.31 $\pm$0.05 \\
1136$-$135 &      &       &   &    1 &  0.07 $\pm$0.05 \\
\hline
\end{tabular}
\end{table}

The component sizes are estimated from the best fits to the observed power spectrum after trying different values of component sizes and parameters of the standard solar wind model
(Manoharan \& Ananthakrishnan 1990; Manoharan et al. 1994;
Gothoskar \& Rao 1999). The free parameters in this model
are the solar wind velocity, random velocity component in the scintillating medium,
axial ratio of the irregularities in the medium, the power-law index of the 
power spectrum of the turbulence of the medium, and  the size of the
scintillating source. The model fitting was done 
only for sources observed with a high signal to  noise ratio. 
Even though a 5 mas difference can be distinguised from the power spectrum, the 
typical inner-scale of the interplanetary medium corresponds
to  an angular scale of 50 mas, which is the typical error in the estimated
angular sizes.

If the structure of the scintillating component is 
simple, then it might be possible to have an idea
of the two-dimensional structure of the source if the direction of the solar wind covers 
a wide range of position angles across the source.
For example, in the case of an elliptical gaussian brightness distribution, 
a variation of the estimated size from a minimum to a maximum should be
observed if the solar wind covers a range of position angles $\ge 90^\circ$
during the different observations of the source.
There are 19 sources showing variation of the estimated  size with the
position angle of the direction of motion of the solar wind. Assuming the
structure to be elliptical, we have estimated the parameters of 4  
sources where we had enough data to represent the ellipse.

The results of our analyses are presented in Table 4, which is 
arranged as follows: column 1: IAU name; column 2: number of observations used  
to estimate the  size  of the scintillating component; column 3:
estimated size of the scintillating component. For
the four sources  whose two-dimensional structure has been determined assuming it to be
elliptical,  the major and minor axes and the position angle of the major axis
are quoted.  The sources which exhibit variation in size, but without enough data to 
determine the position angle of an elliptical structure unambiguously,
the maximum and minimum size are quoted. For the remaining sources, the average size
is listed.
column 4: the  total number of IPS observations;
column 5: the fraction of flux density of the scintillating  component, $\mu$, and the
percentage error in $\mu$.

\section{Hotspots in CSS and larger sources}

The hotspots, which are the high brightness regions in the outer lobes of the
high-luminosity FRII radio sources, indicate the 
Mach disks where the jets terminate. The hotspots usually  subtend small angles
in the radio cores suggesting the high degree of collimation of the jets in
these sources (Bridle \& Perley 1984; Bridle et al. 1994; Fernini et al. 1997).
However, the jet momentum may be spread over a larger area than the cross-section of
the jet itself due to the dentist-drill effect discussed by Scheuer (1982),
where the end of the jet wanders about the leading contact surface drilling
into the external medium at slightly different places at different times.
Recent simulations of 3D supersonic jets suggest that cocoon turbulence
drives the dentist-drill effect (Norman 1996).

In this Section we study some of the properties of the hotspots for CSS as
well as larger objects using the results from our IPS survey and also 
available interferometric observations of hotspots. IPS observations 
enable us to estimate the weighted average of the fraction of flux density in the
scintillating components and their sizes. For sources larger than about 400
mas the scintillating components are usually the hotspots at the outer edges
of the lobes. Since the cores in CSSs are usually weak, especially at low
frequencies, the scintillating components are the hotspots in the lobes.
Over the last few years the sizes of hotspots have also been determined 
reasonably reliably from interferometric observations for
samples of compact steep-spectrum radio sources  using largely VLBI and MERLIN
observations, as well as for the larger objects using the VLA. Although
there is no well-accepted definition of a hotspot (cf. Laing 1989; Perley 1989),
we have used the following empirical definition. The hotspots
are defined to be the brightest features in the lobes located further from
the nucleus than the end of any jet, and in the
presence of more extended diffuse emission these should be brighter by at least a factor
of 4 (cf. Bridle et al. 1994). In the presence of multiple hotspots, only the
primary hotspot has been considered.

%FIG1
\begin{figure*}[t]
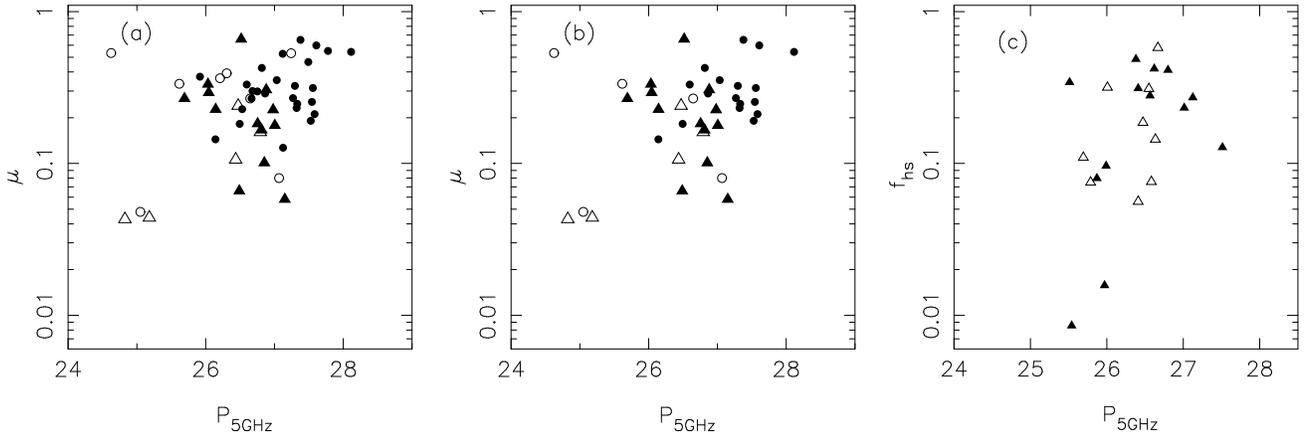

\vspace{-8.0in}
\hbox{
\hbox{
\hspace{-0.3in}
\psfig{figure=9299.f1a}
}
\hspace{-5.0in}
\hbox{
\hspace{-0.7in}
\psfig{figure=9299.f1b}
}
\hspace{-5.0in}
\hbox{
\hspace{-0.7in}
\psfig{figure=9299.f1c}
}
}
\caption{ The scintillation visibility, $\mu$ for sources (a) with LAS $>$ 400 mas, (b) LAS $>$ 1 arcsec 
and (c) the fraction of emission
from the hotspots, f$_{hs}$, estimated from interferometric observations by B94 and F97 
are  plotted against the total radio luminosity 
at 5 GHz. The sample of largely CSS and GPS sources (Table 1) are shown
by open and filled circles for galaxies and quasars respectively, while the larger galaxies and
quasars (Table 2, B94 and F97) are shown by open and filled triangles respectively. 
}
\end{figure*}

In the following subsections we consider the possible dependence of the fraction of flux
density from the hotspot and its size on both radio luminosity and overall projected 
linear size, and discuss possible constraints these might place on models of
evolution of radio sources. We have excluded sources with prominent flat-spectrum
nuclei, and those with a complex or core-jet morphology, and have considered only those 
sources where the scintillations are likely to be produced by the hotspots in the outer lobes. 
In the GPS sources with two outer components, we have assumed that these features 
are likely to be hotspots rather than being the counterparts of flat-spectrum nuclei.
The compact-double structures seen in several GPS sources, lack of variability, and
polarization measurements as in 2134+004 (cf. Stanghellini et al. 1998b) lend some 
support to such an interpretation. 

\subsection{Hotspot prominence and radio luminosity}
The dependence of the prominence of the hotspots on radio luminosity 
was examined about 20 years ago (e.g. Jenkins \& McEllin 1977; Kapahi 1978).
Jenkins \& McEllin reported a strong correlation of the prominence of the 
hotspots with radio luminosity for the well-studied sample of 3CR radio 
sources. 
They defined the hotspots to be features with a size less than about 15 kpc.
Kapahi (1978)  argued that this correlation is possibly due to the
effective resolution being coarser for the higher redshift and hence higher 
luminosity sources. 
We investigate this relationship for the CSS as well as
larger objects using both the IPS and interferometric measurements. 

In Figure 1a and b the scintillation visibility, $\mu$, is plotted against
the total radio luminosity at 5 GHz for the sources from our IPS observations
(Tables 4 and 5) with an LAS $>$400 and 1000 mas respectively. Since the 
effective size of the hotspots estimated from IPS observations can depend on
the relative prominence of the compact and halo components as well as their
separation (cf. Duffet-Smith 1980), we have examined these trends for
sources $>$1000 mas where the blending of the two oppositely-directed hotspots
is minimal. The scintillation visibility, $\mu$, ranges from about 0.04 to 0.7
with most objects in the range of $\sim$0.1 to 0.7. The median value of 
$\mu$ is about 0.3. The luminosity of most objects at 5 GHz lie in the 
range of about 10$^{25.5}$ to 10$^{28}$ W Hz$^{-1}$ sr$^{-1}$, and in 
further discussions
in this paper we confine ourselves largely to objects in this luminosity range. 
The scintillation
visibility $\mu$ - luminosity diagrams for all the sources in Tables 1 and 
2 with S$_{327} >$ 0.5 Jy (Figures 1a and b), show no evidence of a significant 
dependence on radio luminosity. 

We have also examined this relationship using the hotspot flux densities
listed by Bridle et al. (1994, hereinafter referred to B94) and  Fernini et al. (1993, 1997, 
hereinafter referred to as F97), whose sources have
a similar luminosity to those of our samples and have been observed with resolution of
about a few hundred mas, which is comparable to our IPS cut-off size.
A plot of the fraction of the hotspot flux densities, f$_{hs}$ = (S$_{h1}$+S$_{h2}$)/S$_{total}$,
from the two lobes against the total radio luminosity at 5 GHz for the sources from B94 and F97
are presented in Figure 1c. 
The interferometric measurements also do not show a significant dependence of hotspot
prominence on radio luminosity. 

%FIG2
\begin{figure*}
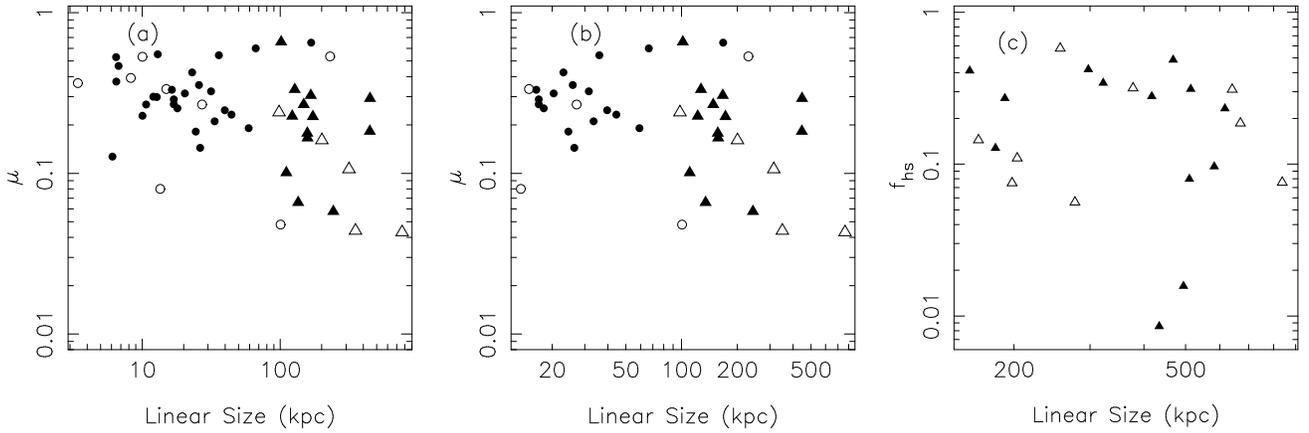

\vspace{-8.0in}
%\hspace{-1.0in}
\hbox{
\hbox{
\hspace{-0.3in}
\psfig{figure=9299.f2a}
}
\hspace{-5.0in}
\hbox{
\hspace{-0.7in}
\psfig{figure=9299.f2b}
}
\hspace{-5.0in}
\hbox{
\hspace{-0.7in}
\psfig{figure=9299.f2c}
}
}
\caption{ The scintillation visibility, $\mu$ for sources (a) with LAS $>$ 400 mas, (b) LAS $>$ 1 arcsec
and (c) the fraction of emission
from the hotspots, f$_{hs}$, estimated from interferometric observations by B94 and F97
are  plotted against the projected linear size. 
The symbols have the same meaning as in Figure 1.}
\end{figure*}

\subsection{Hotspot prominence and linear size}
We have presented the $\mu$-linear size diagram for the
IPS samples described in the earlier section in Figures 2a and b, and the f$_{hs}$-linear size 
diagram for the B94 and  F97 sources in Figure 2c. 
The Spearman rank correlation coefficient for the IPS sample of sources with
LAS $>$ 1000mas is $-$0.32, compared to $-$0.15 for the B94 and F97 sources.
Again, we find no evidence of a significant dependence
of either $\mu$ or f$_{hs}$ on linear size for these high-luminosity objects. 
There is, at best, a weak anticorrelation.

A relation between the prominence of hotspots with source size,
and  the total luminosity  can in principle provide constraints on
the models of  evolution of radio sources. The
luminosity of the hotspot depends on the pressure in the hotspot and the
size of the hotspot.
Many models in the literature assume that the contribution
of the hotspot to the total luminosity is negligible
(Kaiser et al. 1997;
Blundell et al. 1999) or do not consider the evolution of
the hotspot independently (Chyzy 1997; Begelman 1996).
The self-similar models assume that the
pressure in the head of the jet, which is essentially the hotspot, scales with the
mean cocoon pressure (Kaiser et al. 1997;
Begelman 1996) by means of adjusting the size of the working surface.
These models predict that the cocoon luminosity decreases as the source ages.
This would suggest, in the light of our data, that the hotspot luminosity
also decreases as the source grows old. Non-relativistic numerical simulations
have been attempted to understand the structures of the hotspots
(Wilson \& Scheuer 1983; Smith et al. 1985; \cite{NormanBalsara}). 
Although these simulations reveal dynamically varying structures of the hotspots,
the luminosity evolution needs to be studied. The simulations
involving relativistic electron transport with the 3D MHD simulations
of jets (Jones et al. 1999a; Tregillis et al. 1999; Jones et al. 1999b)
might provide better insight into the evolution of the hotspots.

%FIG3
\begin{figure*}
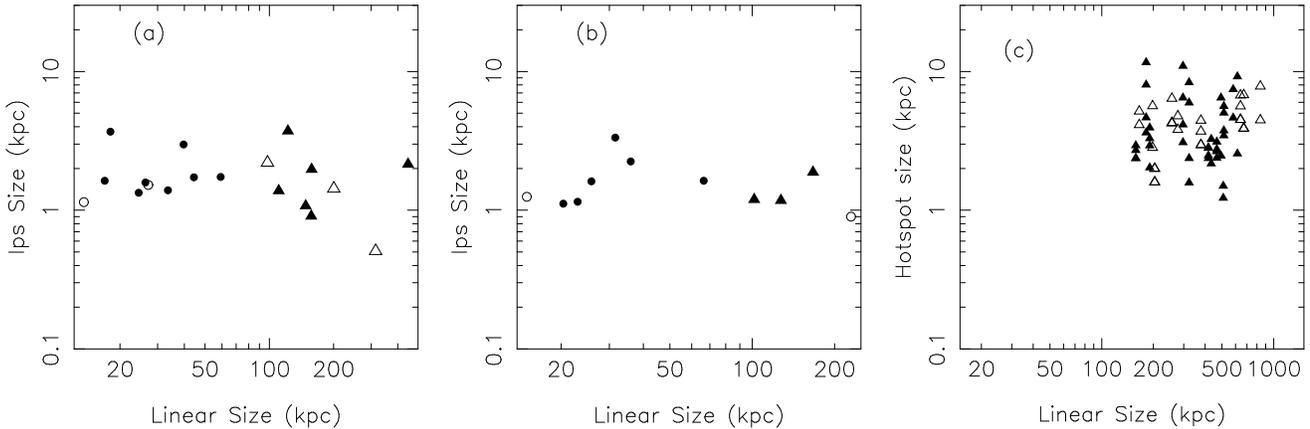

\vspace{-8.0in}
\hbox{
\hbox{
\hspace{-0.3in}
\psfig{figure=9299.f3a}
}
\hspace{-5.0in}
\hbox{
\hspace{-0.7in}
\psfig{figure=9299.f3b}
}
\hspace{-5.0in}
\hbox{
\hspace{-0.7in}
\psfig{figure=9299.f3c}
}
}
\caption{The linear sizes of the scintillating components 
for all the objects in Tables 1 and 2 with an LAS $>$ 1 arcsec  are plotted against the
total projected linear size in panel (a) and for those with $\mu >$ 0.3 in panel (b). In 
panel (c) the sizes of the hotspots from B94 and F97 
are plotted against the projected
linear sizes of these objects. The symbols have the same meaning as in Figure 1.
}
\end{figure*}

\subsection{Sizes of hotspots and collimation of radio jets}
A study of the variation of the size of the hotspots on the overall linear size of the
objects could provide useful clues on the collimation of radio jets. A plot of the
size of the scintillating components for sources larger that 1000 mas, against the overall
linear size of the source shows no significant dependence of the hotspot size on the
overall size of the object (Figure 3a). To examine any effect of contamination by diffuse
emission around the hotspots (cf. Hewish \& Readhead 1976; Duffet-Smith 1980) 
we have confined ourselves to objects above 1000 mas, and
have also considered separately objects with $\mu >$ 0.3 (Figure 3b). We again find no
evidence of a significant relationship. 
We also examine this trend using the sizes of hotspots determined by B94 and F97, which
have similar luminosity to our IPS sample (Figure 3c).  
The interferometric
measurements for the B94 and F97 sources also suggest that the hotspot sizes do
not exhibit a significant dependence on the overall linear size.
The hotspot sizes remain nearly constant with a mean value of about 3 kpc although the
hotspot sizes range from about 1 to 10 kpc, implying that the jets have an approximately
constant mean width beyond about 20 kpc.

Jeyakumar \& Saikia (2000a,b) have earlier examined the dependence of
hotspot size on projected linear size for CSS and GPS objects, and find that the
hotspot size increases linearly with the total linear size, suggesting that they
evolve in a self-similar way. By comparing this trend with larger objects observed
with a similar number of resolution elements, they suggested that there is a 
flattening of the relationship beyond about 20 kpc. The plots in Figure 3 where most
of the objects are larger than about 10 kpc are consistent
with this flattening. 

The relationship between the hotspot sizes and the overall size of
the source suggest that the
jets are largely confined. 
The jets could be confined by the ambient pressure whose
density falls with distance from the nucleus for sources less than about 20 kpc, while
for larger scales the jets could be possibly magnetically confined. Numerical simulations
of the propagation of jets also show that the hotspot sizes do not tend to increase 
with linear size as the jets propagate outwards beyond a certain distance 
(Sanders 1983; Wilson \& Falle 1985). 

The recollimation of the jet can occur if  the jet pressure falls more rapidly than
the ambient pressure. In such a scenario one would expect  recollimation to occur
at a distance where the jet pressure falls below the ambient pressure.
In the model of Sanders (1983) applied to the jet in NGC 315,
the reconfinement of the jet is accompanied by  conical shocks which heat the jet
causing it to reexpand. In this scenario, the reconfinement region beyond about
20 kpc, where the mean hotspot width has an approximately  constant value of $\sim$3 kpc, requires
a high ambient pressure, comparable to or larger than the jet pressure,
beyond the CSS stage.
On the large scales, where the ambient pressure may not be sufficient for confinement,
the jet could be possibly held together by its own toroidal magnetic field
(cf. Begelman et al. 1984).
Magnetic collimation provides
a natural explanation of the observed trend of a nearly constant width in the large sources.
The mechanism of
self-collimation by current-carrying jets has been examined by Appl \& Camenzind
(1992, 1993a,b), and the development of
Kelvin-Helmholtz and current-driven instabilities in these relativistic MHD jets
have been studied by Appl (1996).
In this scenario, the jet is initially pressure confined and
becomes self-collimated by the magetic pressure when the ambient pressure drops
below the jet pressure. The width of the jet in the pressure confined regime
is determined by the ambient pressure alone, where the jet current is shielded
by the surface currents (Appl \& Camenzind 1992). The increase in jet width with source size in this phase is
due to the decrease in ambient pressure. Self-collimation becomes important when the
ambient pressure falls just below the jet pressure. At this point,
the current, I,  is estimated to be $\approx$10$^{17}$P$^{1/2}$R$_j$
amp where P is in units of 10$^{-12}$ dyn cm$^{-2}$ and the
radius of the jet R$_j$ is in kpc (Appl \& Camenzind 1992).
For a hotspot radius of  1.5 kpc, which is the mean value for our large sources,  and
typical jet pressure of $\sim$$10^{-10}$ dyn cm$^{-2}$  the current required to be carried
by the jet is about 10$^{18}$ amp.   The variation of the sizes of the hotspots 
is consistent with pressure confinement in the
CSS phase with the ambient pressure falling with distance from the nucleus,
while at larger distances from the nucleus where the ambient pressure has fallen
below the jet pressure, the jet could be possibly magnetically confined.

%FIG4
\begin{figure}
\vspace{-7.6in}
\hbox{
\hspace{-0.35in}
\psfig{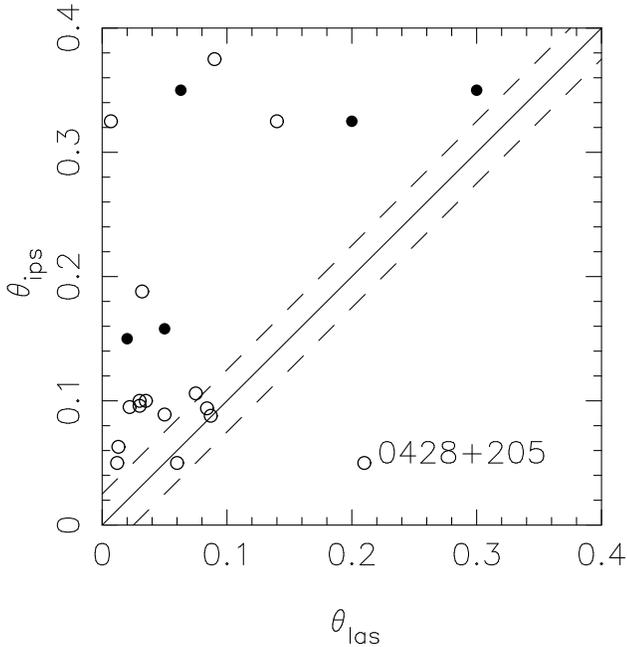}
}
\caption {The size of the scintillating component, $\theta_{IPS}$, is plotted against
the LAS estimated from VLBI observations for CSS and GPS sources whose VLBI structure 
$<$ 400 mas. The symbols are
as defined in Figure 1. The source 0428+205 has been discussed in the text.}
\end{figure}

\section{Extended emission or milli halos around CSSs}

The sources whose overall separation of the components 
are less than the IPS cut-off size 
scintillate as a single source. A comparison of the sizes of the 
scintillating components with the VLBI-scale structure should enable
us to infer the existence of milli-halos on the scale of hundreds of
milliarcsec which may have been resolved out in the high-frequency
VLBI observations, which are usually between 1.4 and 5 GHz. The extended emission 
with steeper spectrum may also be more prominent at low frequencies. 

In Figure 4, we plot the sizes of the scintillating components 
inferred from the IPS observations against the largest angular size
estimated from the VLBI images for all objects with an overall angular
separation less than about 400 mas. There are 21 objects which satisfy
this criterion, and their properties can be found in Tables 1 and 2.

The sizes of the scintillating components estimated from the IPS observations
tend to be larger than the interferometric measurements, except for the
source 0428+205.  Here, the IPS size is significantly smaller than the LAS,
which is defined by the weak extended emission towards the north-west.
The peak brightness of this component is about a factor of 37 weaker than 
the dominant component seen in the VLBI image at $\lambda$18 cm.
The size of the dominant component, whose peak brightness in the
VLBI image is 1476 mJy/beam (Dallacasa et al. 1995) and which is likely
to contribute to most of the scintillating power, is similar to the IPS size.

Since the uncertainty in the IPS size could be as large as about 50 mas, 
we identify only those sources whose difference between the IPS and
VLBI sizes to be at least about 3 times 50 mas to be strong candidates
for milli halos. These include the sources 1127$-$145, 1341+144, 1354$-$174,
1518+047 and 1543+005. Those whose difference lies between 2 and 3 times
are 1442+101, 2247+140 and 2353+154, and these are possible candidates for
millihalos. These 8 sources are described briefly below. The IPS sizes of 
the remaining 12 sources with LAS $\lapp$400 mas, namely 0019$-$000, 0320+053, 
0742$+$103, 0752$+$342, 0941$-$080, 1117$+$146, 1245$-$197, 1345$+$125, 
2008$-$068, 2128+048, 2147$+$145 and 2210$+$016, are consistent with the VLBI sizes.  

There appears to be evidence of extended emission around some of the CSSs on scales
few times larger than the known size of the sources. We refer to these
structures as milli-halos. 
These low-frequency halos are expected to have steeper radio spectra and could
be contributing less to the total flux density at high frequencies; and 
could also have been resolved out in the interferometric maps. There is often
evidence of missing flux density in the VLBI images, but evidence of these
halos from the integrated spectra is difficult to establish because the 
low-frequency integrated spectra as well as the spectra of individual 
components are not well-determined.
One possible scenario  for such structures could
be earlier periods of activity which is presently fading out due to
radiative and other losses.  Another possibility is that the milli-halo is
composed of relativistic particles which have diffused out from the jet
or the nucleus.  Such milli-halos with sizes of the order of about a kpc
have been seen in a few nearby AGN from high-resolution and high sensitivity
observations (Silver et al. 1998; Carilli et al. 1997).
The sizes of extended emission detected in CSSs by IPS observations
are in the range of about 250 pc to  4 kpc. These could be similar to
the milli-halos seen in the nearby AGNs.

\par
\noindent 
{\bf 1127$-$145: } This source has a core-jet type of structure at 2.32 GHz  
with an overall angular size of about 63 mas (\cite{60}). At higher resolution
the dominant component is resolved into a double-lobed source with an angular
separation between the peaks of emission of about 3.7 mas (Romney et al. 1984; 
Wehrle et al. 1992; Bondi et al. 1996; Fey et al. 1996). 
The LAS of the source estimated from the IPS observations is about 350 mas, 
significantly larger than the known VLBI-scale structure.
\par
\noindent 
{\bf 1341+144:} The source 1341+144 is a possible double with an angular 
separation between the peaks of emission of about 60 mas (Cotton et al. 1989).
The LAS inferred from the IPS observations is about 375 mas.
\par
\noindent 
{\bf 1354$-$174: } It is a highly asymmetric source where the flux density ratio
of the outer components is about 57, and the overall angular size is about 32 mas
(Frey et al. 1997). The IPS size of the source is 188 mas. 
\par
\noindent 
{\bf 1442+101:} We estimate the size of the scintillating component from IPS
observations to be about 150 mas. The VLBI map 
by Gurvits et al. (1994) shows a core and extension resembling a jet, with the 
largest separation of the components being $\sim$15 mas. 
The VLBI map at 18cm by Dallacasa et al.
(1995) shows a core and jet-like extension with the core in the northern end,
contrary to the Gurvits et al. (1994) image. If the core is towards the south,
the possible jet has a very steep spectrum. The LAS of the source from the
VLBI image is about 24 mas, while 
the size estimated from the IPS observations is about 150 mas.
\par
\noindent 
{\bf 1518+047:} Phillips \& Mutel (1981) find the source to be double-lobed at
18cm with an overall LAS of 135 mas. Both the lobes are resolved 
into two components at 5GHz by  Mutel et al. (1985). A recent image by
Dallacasa et al. (1998) also shows the double-lobed structure. The IPS size of the
source is 325 mas. 
\par
\noindent 
{\bf 1543+005:} It is a possible triple source with an overall angular size
of 7 mas (Stanghellini et al. 1999). The size of the scintillating component
is about 325 mas. 
\par
\noindent 
{\bf 2247+140:}  MERLIN observations by Spencer et al. (1989) show that it
is a single source with an angular size of $\sim$200 mas. VLA observation by 
van Breugel et al. (1984)  show that it is resolved into a possible double-lobed 
source. The IPS size of the source is about 325 mas. 
\par
\noindent 
{\bf 2353+154: } Barthel \& Miley (1988) found this source to have a single 
component with an angular size of 50 mas. The IPS size of the source is about 158 mas.

\begin{table*}
\caption{Estimated parameters of 5 GPS sources with one dominant component}
\label{absorption}
\begin{tabular}{l l l l l l c c l l l}
\hline\\
Source & Id & z & $\nu_p^{T}$ & $\nu_p^{C}$ & $S_p^{C}$ &$\theta_{vlbi}^{C}$ & $\theta_{ips}^{C}$ &$\theta_{SSA}$ & \multicolumn{2}{c}{$n_e$ cm$^{-3}$  } \\
 & & & GHz & GHz & Jy & mas & mas & mas & 10pc & 50pc \\ 
\hline
0019$-$000 &  G & 0.305 & 0.7 & 0.8 &  3.1 & $<$   5.0& $<$    50 & 3.35 &   437 &  195
\\
0400+258 &  Q & 2.109 & 1.4 & 1.5 &  1.1 &   0.3 $-$   2.8&   125 & 1.13 &   845 &  378
\\
0428+205 &  G & 0.219 & 1.0 & 1.7 &  3.3 & $\sim$   1.7& $<$    50 & 1.32 &   964 &  431
\\
0711+356 & Q & 1.620 &  1.7 & 1.9 &  1.6 &   0.1 $-$ $<$   6.9&  $-$  & 0.97 &  1084 &  485
\\
0742+103 & &  & 2.0 & 2.1 &  4.6 &   0.4 $-$   1.7& $<$    50 & 1.36 &  1204 &  538
\\
\hline
\end{tabular}
\end{table*}

%FIG5
\begin{figure}[t]
\hbox{
\psfig{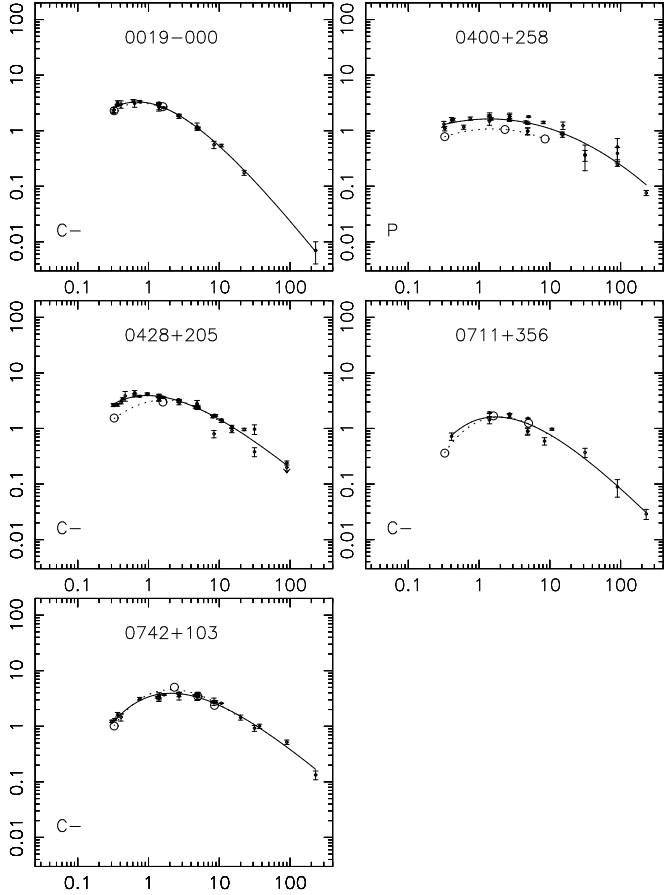}
}
\caption{The integrated spectra and the spectra of the dominant component
from VLBI and IPS observations for GPS sources which have one dominant
component. The x-axis is in units of GHz while the y-axis is in units of Jy. 
The filled circles and the continuous lines represent the
integrated spectrum, while the VLBI and IPS flux densities of the dominant
component are denoted by open circles. The dotted lines denote the fits to
the spectra of the components using these measurements and the integrated
flux densities at higher frequencies, except for 0428+205 where only the
component flux densities have been fitted. The spectra have been fitted 
using the expressions log S = a$_0$+a$_1log\nu$+a$_2(log\nu)^{2}$ for 0428+205
which is marked P, and log S = a$_0$+a$_1log\nu$+a$_2 e^{-log\nu}$ for the 
remaining four sources which are marked C$-$.
}
\end{figure}

\section{ Spectra of scintillating components in GPS sources }
The spectral turnover of GPS sources at low radio frequencies could be either
due to synchrotron self absorption or free-free absorption. In order to
try and distinguish between these two principal processes, we attempted
to determine better the spectra of the compact component using both
IPS observations at low frequencies and VLBI observations at higher
frequencies. Since all the emission with compact structures less than
about 400 mas would contribute to the scintillating power, we have 
confined ourselves to those GPS sources in our IPS sample whose emission
at a higher frequency is dominated by a single compact component. The
peak brightness in any secondary component in the VLBI images is less 
by at least a factor of about 10, and it is reasonable to assume that
the scintillations are almost entirely from the dominant component. 
Five  of the objects in our sample satisfy this criterion. 

We present the integrated spectra as well as that of the dominant
component from the IPS and VLBI observations in Figure 5; and the
results are summarized in Table 6 which is arranged as follows.
Column 1: source name; column 2: optical identification; 
column 3: the redshift; column 4: frequency of the peak in the 
integrated spectrum, $\nu_p^{T}$,  in 
GHz; column 5: frequency of the peak in the spectrum of the dominant 
component, $\nu_p^{C}$  in GHz;
column 6: the flux density of the dominant component at the peak frequency;
columns 7: the range in the VLBI sizes of the prominent components in mas, with
each size being the geometric mean of the major and minor axes. 
Column 8: the size of the IPS component 
in mas; column 9: the expected size of the dominant 
component if the turnover is due to synchrotron self absorption.
The angular size of the self absorbed component can be estimated from
(cf.  Kellermann et al. 1981)

$\theta ~~\approx~~ 13.45 ~~~\nu_p^{-5/4}~~~ S_p^{1/2}~~~ B^{1/4}~~~(1 + z )^{1/4}$

\noindent where $\theta$ is in mas, peak frequency $\nu_p$  in GHz,
the flux density at the peak frequency, $S_p$ is in Jy and the magnetic field B in Gauss. We have assumed a redshift of 1 for the source 0742+103
Column 10 and 11: the electron density of the absorbing medium, if the absorption is due to thermal
free-free absorption. This is given by (cf. Osterbrock 1989)

n$_e^2 ~~\approx~~ 3.05 \times 10^6 ~~~T^{1.35}~~~\nu^{2.1} ~~~({1 \over L})$ cm$^{-6}$

\noindent where $\nu$ is in GHz, T is in units of 10$^4$ K and the size of the absorbing medium L in pc.
We have assumed T=10$^4$ K and have listed the values of n$_e$ for L=10 and 50 pc.

The expected size of the components if the turnover is due to 
synchrotron self absorption ranges from about 1 to 5 mas assuming a 
magnetic field of 10$^{-4}$ Gauss (e.g. Mutel et al. 1985).
The sizes and spectra of the components are generally consistent with synchrotron
self absorption, although models involving free-free absorption are viable
(cf. Bicknell et al. 1997). In the radio source 0400+258 
the total linear separation of the components extends to over 10 mas (Fey \& Charlot 1997),
and could extend to larger sizes with the extended structure appearing somewhat diffuse, although the size of the individual components
range from 0.3 $-$ 2.8 mas.  If the low-frequency turnover is due to free-free absorption the
required densities  of an absorbing medium of thickness 10 and 50 pc
at a temperature of 10$^4$ K,  are listed in Table 6. These values are in the range
of 200 to 1200 cm$^{-3}$. 
Carvalho (1998) suggests that this dense gas might imply the frustration scenario where CSSs 
are as old as the larger FRII sources.  However, estimates of this density could be reduced
as in the model of Bicknell et al. (1997).

Recently Kuncic et al. (1998) considered the effect of induced 
compton scattering (ICS) by the external shell of thermal gas
(Bicknell et al. 1997) to explain the low frequency turnover of the GPS sources.
The Thomson optical depth ($\tau_T$)of the ICS screen ranges from about 0.001 to 0.04 for
the model parameters quoted in Kuncic et al. (1998). 
Using the peak frequency of the components (Table 6), we estimate the
angular sizes of the components to be in range of about 1 to 8 mas
for $\tau_T$ of 0.001 and 7 to 50 mas for $\tau_T$ of 0.04. These values
are consistent with the component sizes listed in Table 6, making it difficult
to distinguish unambiguously between the different processes responsible for the
low-frequency turnover.

\section{Concluding remarks}

We have determined the small-scale structure of samples of 
CSS, GPS and larger sources at 327 MHz using the technique of
IPS with the Ooty Radio Telescope. We have also compiled the
information on the structure of these sources and summarise 
the conclusions based on both our IPS and 
interferometric measurements.

\begin{enumerate}

\item The prominence of the hotspots shows very little
dependence on the luminosity of the radio source over few
decades of luminosity from about 10$^{25.5}$ to 10$^{28}$ 
W Hz$^{-1}$ sr$^{-1}$. The prominence of the hotspots 
also shows no significant dependence on the linear size over about
two decades in linear size. These are based on both the IPS observations
as well as interferometric observations from the literature.

\item The size of the hotspots were shown earlier to increase linearly
with the projected linear size for the CSS and GPS sources
upto about 20 kpc, and then flattens at larger distances by 
considering objects observed with a similar number of resolution
elements (Jeyakumar \& Saikia 2000a,b). The sizes of the hotspots
estimated from the IPS observations as well as the interferometric
observations of hotspots with resolutions comparable to the
IPS cut-off size are consistent with this flattening.

\item We identify candidates for milliarcsec `halos' at 327 MHz in 8
CSS and GPS objects. These could be due to earlier periods of
activity or diffusion of relativistic electrons from the jets and
lobes. VLBI imaging at low-frequencies of these sources would be
useful to clarify the situation.

\item We have determined the spectra of the dominant component
in five GPS objects from both IPS and VLBI observations in order to
distinguish between free-free and synchrotron self absorption.
The turnover in their spectra is consistent with synchrotron 
self absorption although models involving free-free absorption are
viable.
\end{enumerate}

\begin{acknowledgements}
We thank Divya Oberoi, and all the observers at Ooty, particularly Ravi Kumar, Magesh 
and Raghunathan, for their help during the observations, and Pradeep Gothoskar and 
P.K. Manoharan for numerous discussions to clarify the finer points of IPS.
We thank an anonymous referee, Pradeep Gothoskar, V. K. Kulkarni and P. K. Manoharan for their 
valuable comments on the manuscript. We have made use of the
NASA/IPAC Extragalactic Database, operated by the Jet Propulsion Laboratory,
California Institute of Technology under contract with NASA.
\end{acknowledgements}

\end{document}